# Semantic Image Retrieval by Uniting Deep Neural Networks and Cognitive Architectures

Alexey Potapov[1,3], Innokentii Zhdanov[2,3], Oleg Scherbakov[2,3], Nikolai Skorobogatko[2], Hugo Latapie[4], Enzo Fenoglio[4]

[1] SingularityNET Foundation, The Netherlands
[2] Novamente LLC, USA
[3] ITMO University, Kronverkskiy pr. 49, 197101 St. Petersburg, Russia
[4] Chief Technology & Architecture Office, Cisco
pas.aicv@gmail.com, avenger15@yandex.ru, scherbakovolegdk@yandex.ru, nicksk@mail.ru, {hlatapie, efenogli}@cisco.com

**Abstract.** Image and video retrieval by their semantic content has been an important and challenging task for years, because it ultimately requires bridging the symbolic/subsymbolic gap. Recent successes in deep learning enabled detection of objects belonging to many classes greatly outperforming traditional computer vision techniques. However, deep learning solutions capable of executing retrieval queries are still not available. We propose a hybrid solution consisting of a deep neural network for object detection and a cognitive architecture for query execution. Specifically, we use YOLOv2 and OpenCog. Queries allowing the retrieval of video frames containing objects of specified classes and specified spatial arrangement are implemented.

## 1 Introduction

Bridging symbolic/subsymbolic gap (e.g., [22]) is one of difficult problems in artificial general intelligence (AGI) and cognitive architectures (CAs) in particular. This problem has many manifestations in practical tasks. One such task is the semantic image retrieval, which involves both subsymbolic processing of images or videos, and queries defined on a symbolic level describing the semantic content of images to be retrieved. This task is also practically important. One might want to find specific images in a photo collection or a video frame with certain content.

Due to its practical importance, semantic image retrieval has been intensively studied within traditional AI areas. However, conventional computer vision methods were able to recognize not too many classes of objects simultaneously. Thus, many efforts were directed towards bridging the "semantic gap" between low-level image features and high-level concepts in terms of which queries are specified (e.g. [1–3]). Apparently, the semantic gap in computer vision is a particular manifestation of the symbolic/subsymbolic gap.

Deep convolutional neural networks (DCNN) opened the possibility to detect and recognize objects belonging to hundreds and even thousands of different classes.

Moreover, pre-trained DNNs solving this task are readily available, for example, YOLOv2 [4] (You Only Look Once), Deformable R-FCN [5] (Region-based Fully Convolutional Networks), SSD [17] (Single Shot multibox Detector).

However, when we want not just to detect separate objects on images, but to find images with a specified content, modern DCNNs don't provide out-of-the-box solutions (unless being used to perform deep hashing for retrieving images semantically similar to a query image, e.g. [6, 7]). Although there are some successes in purely DCNN-based image understanding including image caption generation [8] and visual question answering [9], such systems don't directly solve the image retrieval task, they are difficult to train and are less flexible in comparison with traditional knowledge-based image understanding systems. Thus, constructing a purely neural system that learns a mapping between visual and linguistic data is still not too practical now, at least, for image retrieval. Moreover, a question whether the purely neural-based approach is optimal for AGI is controversial.

Thus, the most accessible benefit of deep learning in the semantic image retrieval now is object detection and recognition. However, a hard-coded engine for executing a limited set of queries based on detected objects is also not useful enough, and usage of a knowledge-based reasoning is desirable. CAs as modern intelligent systems that usually support knowledge representation and reasoning are underutilized here.

In this paper, we investigate if it is possible to efficiently use CAs, namely, OpenCog, in combination with DCNNs to construct a semantic image retrieval system. Although we report preliminary results achieved without tight integration of the symbolic and subsymbolic components, these results show that even a loose integration provides practical benefits for semantic image retrieval, which is a good testbed for studying the problem of bridging the symbolic/subsymbolic gap.

## 2  Previous Works

**Semantic Image Retrieval**

A common approach to retrieve images with semantic structure through the learning is to train deep models on images with joint labels of several classes which form complex concepts and learn a relationship model that represents the expected spatial relationships among the relevant objects for retrieval of instances of visual situations [10]. For example, if we have labels for person and bicycle with corresponding relation we can train model to a new concept – cyclist. However, such procedure requires exhaustive labeling (including forming of negative examples).

Also there are a plenty of traditional methods such as [11, 12] that use low-level hand-crafted features for image representation along with relatively simple text-clustering techniques. Needless to say that such limited representations lead to poor performance when applied to wide range of image retrieval problems.

**Knowledge-based System for Image Retrieval**

Another approach to image retrieval task is the one based on knowledge manipulating systems. These techniques mostly shift focus from the quality of image representation to consistent work with complex structure of semantic relations between concepts. Also such approach provides useful tools for construction of languages for visual programming.

Some of the methods [13] use knowledge parsers along with popular knowledge datasets such as ConceptNet or WordNet to improve retrieval accuracy. Some others [18] use conditional random field models defined over a scene graph representing the query for semantic image retrieval. Here, the scene graph captures the detailed semantics of visual scenes by explicitly modeling objects, attributes of objects, and relationships between objects and assumes existence of rich concept graphs which are usually immutable. So it makes such methods non-flexible and hardly extensible. Some of the methods [14] use self-organizing maps (SOM) for concerted high-level semantic and low-level visual features analysis. Obviously these methods have limitations caused by expressive power of SOM.

Apparently, both expressive image representations and structured knowledge are needed for semantic image retrieval.

## 3 Proposed System

**Object Detector**

State-of-the-art DCNN object detectors can be divided into two groups: region proposal-based methods and proposal-free methods. Proposal-based methods like R-FCN [5, 20] are two-stage detectors that start generating a set of candidate bounding boxes (BBs), and then focus on processing each candidate. Proposal-free methods like YOLO [4, 19] are single-stage detectors that consider detection a regression problem, use a single ConvNet and run once on the entire image.

We considered two deep convolutional neural networks, to detect and recognize objects: YOLOv2 [4, 19] which offers a competitive speed and Deformable R-FCN [5] which offers a good trade-off between detection efficiency and accuracy. Both networks were trained on the same MS COCO [15] dataset with 80 objects category.

The detector is one of the key components of the system, so it was important to compare the performance of both networks with our data sets. For this purpose, a video was mounted and synchronized to show the output frames from YOLO and Deformable R-FCN.

As a result of the comparison, we made the following observations (see Figure 1):
- Deformable R-FCN network marks out the same detected object at once with many frames (basically same class). See Figure 1(a).
- In general, the Deformable R-FCN classification has fewer errors than YOLO, at the cost of fewer detected objects. See Figure 1(b).

- The YOLO network detects more different objects that are interesting for the semantic video frame retrieval than the Deformable R-FCN; these objects can be a part of the interior, an element of human clothing, etc. See Figure 1(c).

Additionally, the detection threshold of YOLO can be changed to display objects detected with a higher (lower) confidence by increasing (decreasing) the parameter "threshold", but the number of interesting objects will be affected accordingly. In general, the YOLO network is more preferable for extracting data about objects on video. However, in some cases, this network cannot be used because of the large number of classification errors.

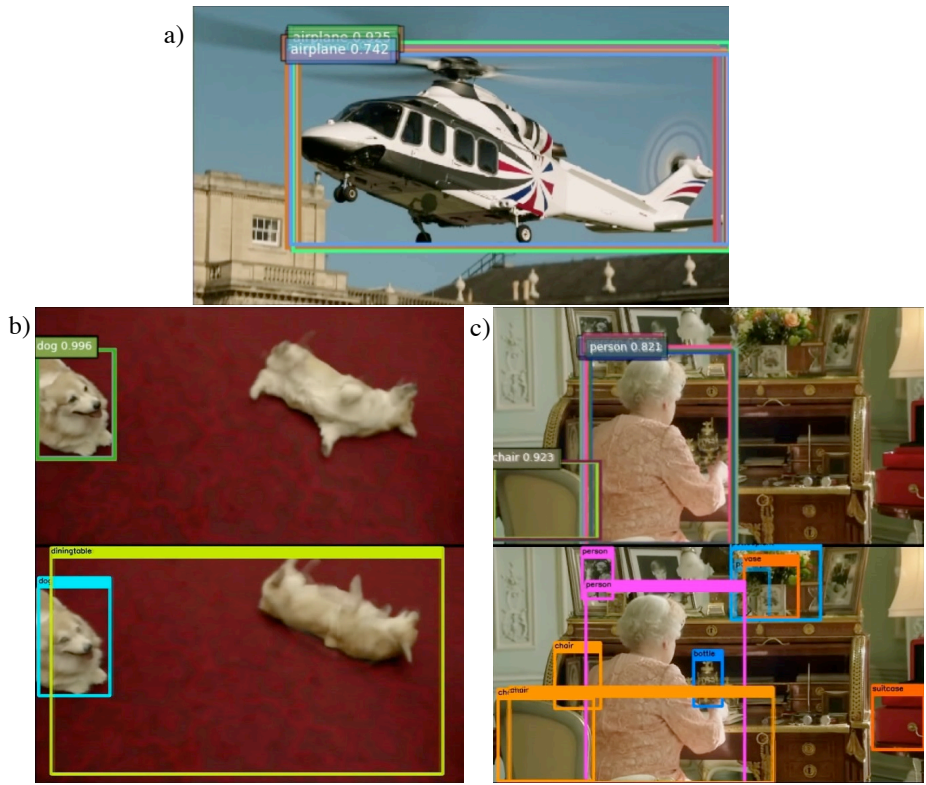

**Fig. 1.** Sample of Deformable R-FCN detection; (b,c): the upper image corresponds to Deformable R-FCN, the lower image corresponds to YOLO

**Implementation of Queries in OpenCog**

OpenCog is a cognitive architecture built on the top of a hypergraph-based knowledge representation and a powerful inference engine. The container for these hypergraphs is called *AtomSpace*. We will use just a small part of its functionality, addressing the interested reader to the detailed description referenced in [16]. For our current purposes, it is enough to treat its knowledge representation as an ordinary graph (except Bind link used in the inference), which is filled with information about detected bounding boxes including their coordinates and labels.

Figure 2 shows an example of a fragment of this graph describing two bounding boxes belonging to one frame.

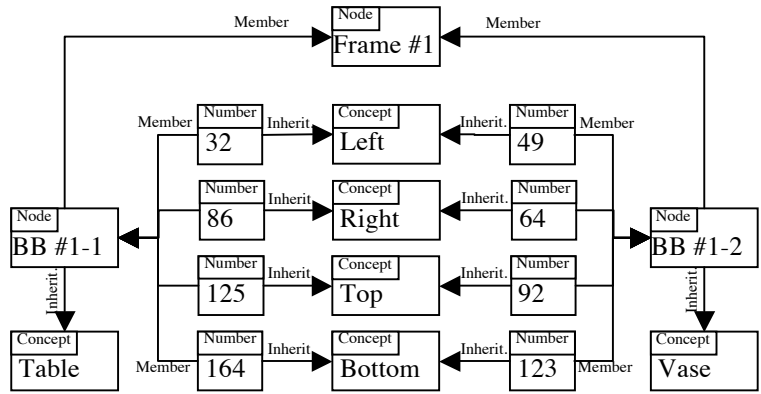

**Fig. 2.** Description of two extracted bounded boxes in one frame as a part of AtomSpace

One way to perform the inference in OpenCog is through "Pattern Matching", i.e. matching a template (with variable nodes) sub-graph (which is also stored in AtomSpace) against the rest of AtomSpace. The matching result can also be placed into AtomSpace via the activation of a special type of links, e.g. Bind link.

Consider the Bind link shown in a slightly simplified form in Figure 3.

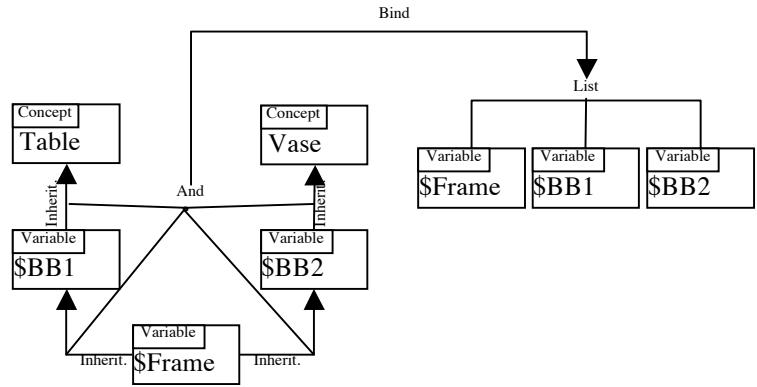

**Fig. 3.** Example of Bind link for retrieving a sub-graph corresponding to two specific objects presented in one frame

Left part of this link can be matched against a sub-graph of the graph presented in Figure 2. Thus, by activating this link one can retrieve a sub-graph containing the List link uniting the BB#1-1, BB#1-2 and Frame#1 nodes.

Thus, such Bind link can be used to retrieve frames containing required objects. We use Python API to AtomSpace to synthesize Bind links for specified objects, although this could be done inside AtomSpace (e.g. by Bind links over Bind links).

More types of queries can be implemented using coordinates of bounding boxes and additional types of links. Coordinates of bounding boxes can be bound with variables in the same way as it was done above for BB nodes. These coordinates can be compared using GreaterThan links, and one can use And link to require several conditions to be hold simultaneously. For example, the following code in *Atomese* (a programming language to describe the content of AtomSpace in the text form) specifies a template graph (similar to those shown in the previous figures) that can be used to find bounding boxes in the same frame, one of which is left to another one.

```
AndLink
   MemberLink
      InheritanceLink
         VariableNode "$Left1"
         Node "Left"
      VariableNode "$BB1"
   MemberLink
      InheritanceLink
         VariableNode "$Right2"
         Node "Left"
      VariableNode "$BB2"
   MemberLink
      VariableNode "$BB1"
      VariableNode "$Frame"
   MemberLink
      VariableNode "$BB2"
      VariableNode "$Frame"
  GreaterThanLink
     VariableNode "$Left1"
     VariableNode "$Right2"
```

To utilize modularity, we defined a number of Bind links, which insert intermediate inference results into AtomSpace. For example, the code shown above was used in a Bind link with the following resultant:

```
EvaluationLink
   PredicateNode "RightTo"
      ListLink
         VariableNode "$BB1"
         VariableNode "$BB2"
```

This Atomese code means that the predicate "RightTo" evaluates to true for a pair of nodes bound to the Variable nodes "$BB1" and "$BB2". It is not necessary to represent the resultant as a predicate. What is needed is to create a sub-graph which contains the necessary information and which can further be pattern-matched. Nevertheless, predicates seem quite natural here.

Similarly, one can define such predicates, which will be true for intersecting BBs, or pairs of BBs, one of which is inside another one, or on the top of it, etc. Thus, one can implement such queries as "a vase on a table" or "a painting with a person".

It should be noted that OpenCog supports non-binary truth values, although we don't utilize them in our current implementation, but they should be useful to describe soft versions of spatial relations. Such truth values can also be combined with confidence values assigned to labels by the detector.

With such intermediate conclusions, one-step pattern matching will not be able to find sub-graphs corresponding to queries of interest, e.g. "a vase on a table", without invoking Bind links calculating truth values of helper predicates. OpenCog has to main mechanisms for chaining inference steps, namely, the forward chaining and the backward chaining. The forward chaining starts with the available data and iteratively applies Bind links to fill AtomSpace with resultants. In our task, the backward chaining is more suitable. It starts with a sub-graph (query) of interest and goes backward to find Bind links which can help to infer this sub-graph.

## 4  Experiments

We conducted experiments with some video sequences to validate our approach and test the constructed system. Different queries for retrieving video frames containing specified objects in certain relative locations were executed. Such queries as 'a person inside a car' or 'a person with a bag' were successfully tested. Figure 4 shows some examples of successfully retrieved frames from different videos.

The following queries were used: 'a person inside a car', 'a person left to a car', 'a person with a tie', 'a person with a backpack', and the corresponding bounding boxes are shown in Figure 4. Similar queries can be executed for arbitrary pairs of objects recognizable by the DCNN. Queries involving more than two objects can also be added, but it has not been done yet.

Of course, our image retrieval system can fail in some cases. These failures can be due to incorrect object recognition or by imprecise or not supposed sizes and positions of BBs. Figure 5 shows an example of incorrectly retrieved video frame, because of the recognition error. Figure 6 shows two examples of incorrectly constructed bounding boxes retrieved with the use of unnatural queries. Figure 7 shows another example, for which one can argue that the bounding boxes are not that bad, but 'vase' BB appears to be inside 'flowers' BB. As a result, such frames will not be retrieved by a normal query.

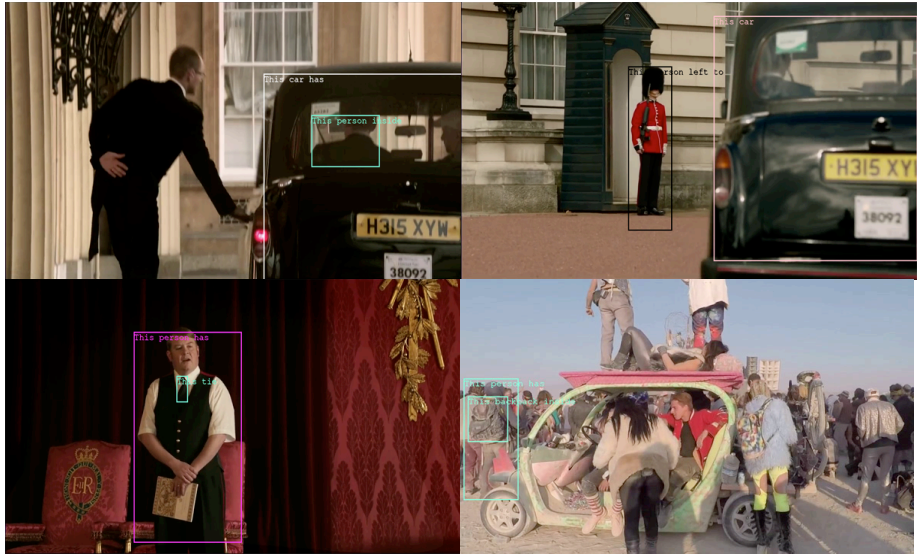

**Fig. 4**. Examples of successfully retrieved video frames

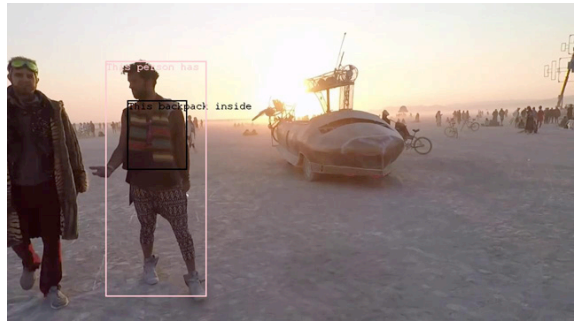

**Fig. 5.** Incorrect retrieval of 'a person with a backpack'

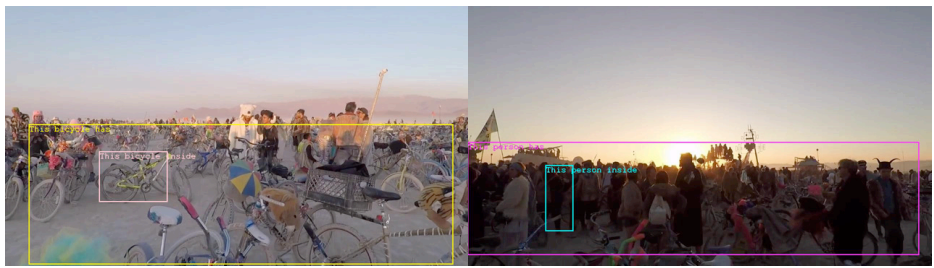

**Fig. 6.** Examples of video frames with incorrect bounding boxes

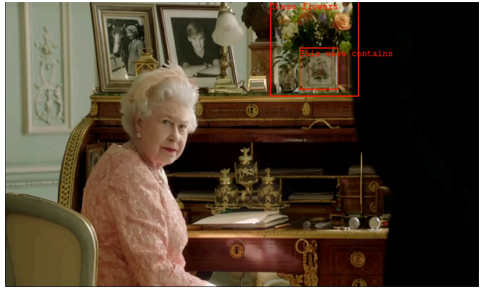

**Fig. 7.** Example of bounding boxes with not supposed arrangement

## 4 Conclusion

In this paper, we have proposed an approach to semantic image retrieval based on integration of DCNNs for object detection and cognitive architectures for semantic analysis and query execution to utilize the power of DNN-based image analysis and flexibility and compositionality of knowledge representation and reasoning in cognitive architectures.

We developed a first version of such system based on YOLOv2 network and OpenCog. We implemented functionality in Atomese language to support queries for retrieving video frames containing specified objects in specified relative spatial positions.

Our results show that this approach is quite practical, and it can be considerably extended in future:

1. One can utilize imprecise probabilities supported by OpenCog to perform probabilistic querying.
2. Language understanding capabilities of OpenCog can be used to create natural language interface for specifying queries.
3. Richer set of queries can be implemented, in particular, to describe events (e.g. approaching of one object to another).
4. Pattern Miner module of OpenCog can be used to automatically create new useful elements of knowledge representation. For example, we may want to recognize visual analogies taking advantage of the "conceptual slippage" in the sense of Hofstadter [21] in which roles defining a situation can be fluidly filled by concepts semantically related to the query and the concepts used in creating the analogies can be considered realization of statistically emergent active symbols formed in the AtomSpace.
5. Detection of possibly incorrectly detected objects or wrong bounding boxes for them using mined patterns in relations between BBs.
6. Events can be handled either on the cognitive level, or with the use of DNNs, or both.

# References

1. Smeulders, A.W., Worring, M., Santini, S., Gupta, A., Jain, R.: Content-Based Image Retrieval at the End of the Early Years. IEEE Trans. Pattern Analysis and Machine Intelligence, vol. 22, no. 12, pp. 1349–1380 (2000)
2. Thanga Ramya, S., Rangarajan, P.: Knowledge Based Methods For Video Data Retrieval. International Journal of Computer Science & Information Technology (IJCSIT), vol. 3, no. 5, p. 165–172 (2011)
3. Manzoor, U., Balubaid, M.A., Zafar, B., Umar, H., Khan, M.Sh.: Semantic Image Retrieval: An Ontology Based Approach. (IJARAI) International Journal of Advanced Research in Artificial Intelligence, vol. 4, no.4 (2015)
4. Redmon, J., Farhadi, A.: YOLO9000: Better, Faster, Stronger. arXiv:1612.08242 [cs.CV] (2016)
5. Dai, J., Qi, H., Xiong, Y., Li, Y., Zhang, G., Hu, H., Wei, Y.: Deformable Convolutional Networks. arXiv:1703.06211 [cs.CV] (2017)
6. Liu, H., Wang, R., Shan, Sh., Chen, X.: Deep Supervised Hashing for Fast Image Retrieval. IEEE Conference on Computer Vision and Pattern Recognition (CVPR), pp. 2064–2072 (2016)
7. Varga, D., Szirányi, T.: Fast content-based image retrieval using convolutional neural network and hash function. IEEE International Conference on Systems, Man, and Cybernetics (SMC) (2016)
8. Karpathy, A., Li, F.-F.: Deep visual-semantic alignments for generating image descriptions. CoRR, abs/1412.2306 (2014)
9. Gordon, D., Kembhavi, A., Rastegari, M., Redmon, J., Fox, D., Farhadi, A.: IQA: Visual Question Answering in Interactive Environments. arXiv:1712.03316 [cs.CV] (2017)
10. Quinn, M, Conser, E, Witte, J., Mitchell, M.: Semantic Image Retrieval via Active Grounding of Visual Situations. arXiv:1711.00088 [cs.CV] (2017)
11. Celebi, E., Alpkocak, A.: Semantic Image Retrieval and Auto-Annotation by Converting Keyword Space to Image Space. 12th International IEEE Multi-Media Modelling Conference Proceedings (2006)
12. Singh, N., Dubey S., Dixit P., Gupta, J.: Semantic image retrieval using multiple features. Natarajan Meghanathan, et al. (Eds): SIPM, FCST, ITCA, WSE, ACSIT, CS & IT 06, pp. 277–284 (2012)
13. Chen, H., Trouve, A., Murakami, K. J., Fukuda, A.: Semantic image retrieval for complex queries using a knowledge parser. Multimedia Tools and Applications, pp. 1-19. DOI: 10.1007/s11042-017-4932-2 (2017)
14. Subramanyam, V., RallabandiSett, R.: Knowledge-based image retrieval system. Knowledge-based systems, vol. 21, iss. 2, pp. 89-100. (2008)
15. Lin, T.-Y., Maire, M., Belongie, S., Hays, J. , Perona, P., Ramanan, D., Dollar, P., Zitnick, C. L.: Microsoft COCO: Common objects in context. arXiv preprint arXiv:1405.0312 (2014)
16. Goertzel, B., Pennachin,C., Geisweiller. G.: Engineering General Intelligence, Part 2: A Path to Advanced AGI Via Embodied Learning and Cognitive Synergy. Atlantis Publishing Corporation (2014)
17. Liu, W., Anguelov, D.. Erhan, D., Szegedy, C., Reed, S.: SSD: Single shot multibox detector. arXiv:1512.02325 (2015)
18. Johnson, J.. Krishna, R., Stark, M., Li, L.-J., Shamma, D., Bernstein, M., Fei-Fei, L.: Image retrieval using scene graphs. IEEE Conference on Computer Vision and Pattern Recognition (CVPR) (2015)
19. Redmon, J., Divvala, S., Girshick, R., Farhadi. A.: You only look once: Unified, real-time object detection. arXiv:1506.02640 (2015)
20. Li, Y., He, K., Sun, J., et al.: R-FCN: Object detection via region based fully convolutional networks. In Advances in Neural Information Processing Systems, pp. 379–387 (2016)
21. Hofstadter, D. R., Mitchell, M.: The Copycat project: A model of mental fluidity and analogy-making. In Advances in Connectionist and Neural Computation Theory, K. Holyoak and J. Barnden, Eds. Ablex Publishing Corporation (1994)
22. Goertzel, B.: Perception Processing for General Intelligence: Bridging the Symbolic/Subsymbolic Gap. Springer: LNCS, vol. 7716 (proc. AGI'12), pp. 79-88 (2012)